# A topological field-effect memristor

Manuel Meyer[1,a], Selena Barragan[1], Sergey Krishtopenko[1,2], Adriana Wolf[1], Monika Emmerling[1], Sebastian Schmid[1], Jean-Baptiste Rodriguez[3], Eric Tournie[3,4] Benoit Jouault[2], Gerald Bastard[1,5], Frederic Teppe[2], Victor Lopez-Richard[6], Ovidiu Lipan[7], Lukas Worschech[1], Sven Höfling[1], and Fabian Hartmann[1,b]

[1]*Julius-Maximilians-Universität Würzburg, Physikalisches Institut and Würzburg-Dresden Cluster of Excellence ct.qmat, Lehrstuhl für Technische Physik, Am Hubland, 97074 Würzburg, Germany*

[2]*Laboratoire Charles Coulomb (L2C), UMR 5221 CNRS-Université de Montpellier, Montpellier, France*

[3]*IES, Université de Montpellier, CNRS, F-34000 Montpellier, France*

[4]*Institut Universitaire de France, F-75005 Paris, France*

[5]*Physics Department, École Normale Supérieure, PSL 24 rue Lhomond, 75005 Paris, France*

[6]*Departamento de Fısica, Universidade Federal de Sao Carlos, 13565-905 Sao Carlos, SP, Brazil*

[7]*Department of Physics, University of Richmond, 28 Westhampton Way, Richmond, Virginia 23173, USA*

**Overcoming the limitations of the von Neumann architecture requires new computational paradigms capable of solving complex problems efficiently. Quantum and neuromorphic computing rely on unconventional materials and device functionalities, yet achieving resilience to imperfections and reliable operation remains a major challenge. This has motivated growing interest in topological materials that provide robust and low-power operation while preserving coherence. However, integrating coherent topological transport with non-volatile memory functionality in a single reconfigurable device has remained challenging. In this work, we demonstrate a topological field-effect memristor based on inverted InAs/GaInSb/InAs trilayer quantum wells operating in the quantum spin Hall regime. The intrinsic floating-gate behavior allows one to reconfigure the transistor functionality into memristive functionality with broad electric-field tunability. Unlike other memristor implementations, one resistance state is governed entirely by dissipationless, coherent transport through helical edge channels, while the other arises from incoherent bulk conduction. By combining electrically tunable coherent and incoherent transport with memory functionality, our device realizes a prototypical topological electronic element that integrates coherent transport and adaptive memristive behavior, paving the way for hybrid quantum-neuromorphic architectures.**

a) Email: manuel.meyer@uni-wuerzburg.de
b) Email: fabian.hartmann@uni-wuerzburg.de

Conventional computing architectures encode information in binary bits high (1 or H) or low (0 or L) and perform operations using Boolean logic. The robustness and fault tolerance of binary computation lies in the large separation between bit values that suppresses the effect of small perturbations[1]. However, this separation also renders binary computation intrinsically energy inefficient. A second limitation - more by choice than fundamentally - is the separation of the central processing unit and memory in the von Neumann architecture. This separation requires a constant data transfer between memory and processing units, leading to sequential computation and performance limitations known as the von Neumann bottleneck, as well as high-power consumption[2,3]. To overcome these limitations,, alternative computational architectures are actively explored[4]. Among them, quantum computing introduces parallelism through superposition and entanglement, leveraging coherence for computation[5]. While quantum computation can be realized on various material platforms with different quantum bit realizations, it demands sufficiently long coherence times for solving complex problems. Thus, fault-tolerance remains the central challenge as coherence is destroyed by environmental factors and disorder[2]. In contrast, neuromorphic computation does not rely on quantum coherence, but its parallel computing ability emerges from the complex interconnection within the network itself[6–8]. While neuromorphic computing can be realized in standard CMOS architectures[9,10], such implementations are energy-inefficient and hence new materials with emergent functionalities are required to solve this issue[11,12]. At first glance, quantum and neuromorphic computing appear mutually exclusive as quantum computing relies on isolation and coherence and neuromorphic computing thrives on interaction with a complex environment. This apparent incompatibility motivates the search for material platforms that can simultaneously preserve coherence and support adaptive electronic behavior. However, a key challenge for both architectures is resilience to perturbations such as disorder, structural imperfections and dissipation, which can disrupt computational outcomes. Therefore, devices that exhibit intrinsic robustness while preserving their functional properties are essential for advancing these technologies.

In recent decades, quantum systems with intrinsic protection have gained significant interest. In particular, topological insulators (TIs), characterized by a non-trivial band topology, offer a compelling route towards low-dissipation and robust information processing even in the presence of perturbations[13]. Time-reversal-invariant two-dimensional (2D) TIs, also known as quantum spin Hall insulators (QSHIs), support protected coherent edge states that are dissipationless and spin-polarized even in the presence of disorder[14–17]. This raises the intriguing possibility of combining (quantum) coherence and neuromorphic adaptability within a single material platform[18–20], potentially reconciling these two seemingly incompatible computational schemes. To explore this possibility, a material system must support coherence-preserving transport while also enabling volatile or non-volatile adaptive behavior characteristic of neuromorphic devices.

Inverted InAs/(Ga,In)Sb-based quantum wells (QWs) are a tunable and scalable III–V semiconductor platform[21,22] that support the QSHI state through band inversion and strong spin-orbit coupling[23,24]. Moreover, these QWs in its symmetrical three-layer configuration possess large topological gaps up to ~50 meV[25,26] with the perspectives of the observation of coherent transport via the topological edge states at elevated temperatures[27–29]. A recent observation of the QSHI state in inverted InAs/GaInSb/InAs trilayer quantum wells (TQWs) at 60 K[30] makes them a promising platform for implementing functional devices that harness topologically protected edge states for electronic applications even at higher temperatures.

Here, we report on the experimental realization of a topological field-effect memristor based on inverted InAs/GaInSb/InAs TQWs, combining coherent helical edge transport with memristive functionality. Our device exploits the broken-gap band alignment between InAs and (Ga,In)Sb, in which appropriate quantum well thicknesses and mole fractions give rise to a band-inverted regime with a topological QSHI state. By harnessing the intrinsic floating-gate behavior of the gate dielectric, we fabricate a topological memristor broadly tunable by electric fields, which demonstrates memristive switching between a bulk-dominated (incoherent) state and a helical edge-dominated (coherent) state leveraging the topological protection. This realization uniquely combines coherent helical edge transport with memristive switching behavior, establishing a new platform for neuromorphic networks based on topological materials.

**Device architecture**

The topological field-effect memristor is based on an InAs/Ga$_{0.68}$In$_{0.32}$Sb/InAs TQW heterostructure, grown on (001) AlSb buffer and engineered to realize the QSHI state. Figure 1 summarizes the device architecture and illustrates its possible dual operation as a field-effect transistor and memristor. As shown in Fig. 1(a), the TQW exhibits an inverted band-gap energy of about 27 meV opened between the two electron subbands (E1 and E2). The first hole subband (H1) lies approximately 20 meV above the E2 band. The relatively large band-gap energy enables the high-temperature observation of the QSHI state[30]. Figs. 1(b) and (c) illustrate a standard transistor configuration, showing a schematic layer structure and an exemplary optical image of a large Hall bar device. In the transistor configuration, a constant bias voltage is applied to the drain contact and the relative position of the Fermi energy ($E_F$) is controlled electrically via the top-gate voltage. The source is on the common ground. By shorting the drain and top-gate electrodes (Figs. 1(d) and (e)), the same device can operate as a memristor, where the floating gate functionality emerges intrinsically[31] due to charge trapping at the semiconductor/oxide interface[28,32]. The oxide used for this device is a superlattice of SiO$_2$/SiN (5x 10 nm/ 10 nm and ending with an additional 10 nm SiO$_2$ layer). Both configurations share a six-terminal Hall bar geometry with a channel width of W = 20 μm and contact separation length between the voltage probes of L = 10 μm (see optical images in panels (c) and (e) for an exemplary device). An additional back-gate voltage is applied to independently manipulate the Fermi energy and the perpendicular electric field utilizing this dual-gating approach[21], providing full electrostatic control of the device. This dual-mode architecture allows direct comparison of transistor and memristor transport within the same topological device. All transport measurements were performed at T = 4.2 K in the dark.

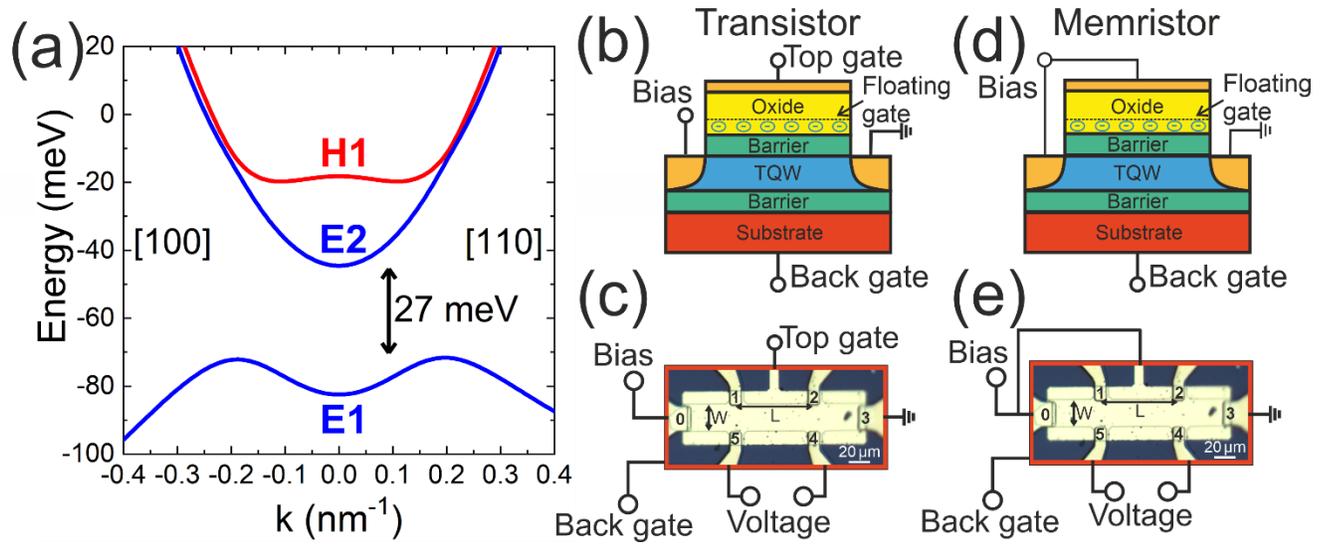

**Fig. 1. Device architecture and dual-mode operation as field-effect transistor and memristor.** (a) Calculated band dispersion of the sample with a topological non-trivial band gap of approximately 27 meV. The gap is formed between the two electron-like (E1 and E2) bands. (b),(c) Transistor configuration: Bias and top-gate voltages are applied independently and via the back-gate voltage, the electric field can additionally be adjusted. (d),(e) Memristor configuration: bias and top-gate voltages are shorted to form a memristor device. Via the back-gate voltage, the electric field can be controlled leading to a three-terminal memristor.

We now examine how topological protection manifests in the field-effect transistor configuration before turning to the memristive operation.

**Quantum spin Hall insulator in field-effect transistor configuration**

To demonstrate how topological protection can enable adaptive electronic behavior, we first examine the device in a field-effect transistor configuration as was shown in Figs. 1(b) and 1(c). The device exhibits tunability via electric fields applied through both the top and back gates. Figures 2(a) and (b) show the resistance as a function of top-gate voltage for two exemplary back-gate voltages $V_{BG}$ = -6 V and -9 V, respectively. The top-gate voltage is swept from $V_{TG}$ = +10 V to -10V (in black, down-sweep) and back to +10 V (in red, up-sweep). In both sweep directions, a prominent resistance peak appears when the Fermi energy lies within the topological-insulating gap. On the right- and left-hand sides of each peak, the Fermi energy lies in the conduction and valence bands, respectively. The insets schematically illustrate the position of the Fermi energy for both sweep directions. A pronounced hysteresis (ΔV) appears between the up and down sweeps. It originates from charge accumulation (ΔQ) at the oxide-semiconductor interface, which acts as an intrinsic floating gate. The relative shift of the peak is governed by the difference in localized charges and the effective capacitive coupling $C_{eff}$: $\Delta V = \Delta Q/C_{eff}$. As seen in Fig. 2(a) for a back-gate voltage of -6 V, both peaks are on the negative top-gate voltage side while for a back-gate voltage - 9 V (see Fig. 2(b)), the peak for the down-sweep direction can be shifted to the positive top-gate voltage region. Hence, for a specific back-gate voltage and thus electric field, the topological gap can be tuned to zero top-gate voltage. Figs. 2(c) and 2(d) show the extracted peak resistance values and their corresponding positions as a function of back-gate voltage. As $V_{BG}$ decreases, the peak resistance increases systematically. Since transport in the band-gap region is mediated entirely by helical edge channels[30], this trend reflects a reduction in the phase coherence length[33]. Concurrently, the peak position shifts towards more positive top-gate voltages, eventually crossing zero at $V_{BG} \approx -7$ V (as seen in Fig. 2(d)).

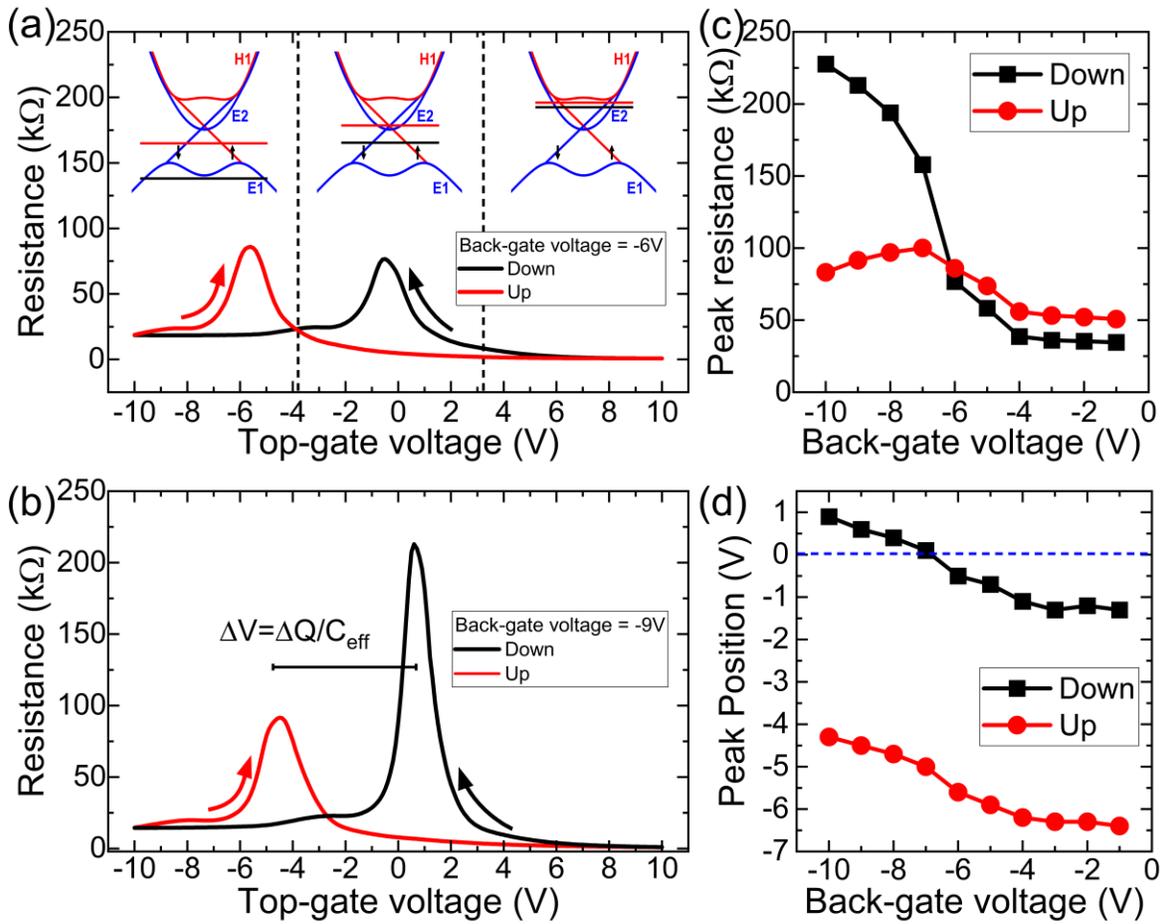

**Fig. 2. Electric field dependence of resistance in transistor configuration.** (a) Resistance as a function of top-gate voltage between +10 V and -10 V (down-sweep, black; up-sweep, red) for a back-gate voltage of -6 V. For positive top-gate voltages, the Fermi energy is in the conduction band, and the dominant charge carrier type is n-type. After the gap region, the Fermi energy is in the valence band, and the transport is mainly p-type. Due to charge accumulation, the up- and down-sweep directions show pronounced hysteresis (ΔV). (b) For a back-gate voltage of -9 V, the position of the gap can be set to positive top-gate voltages. Extracted peak resistance values $R_{max}$ are shown in (c) and their positions in (d) for the up- and down-sweep directions, respectively. At a back-gate voltage of -7 V, the peak position in the down-sweep direction coincides with zero top-gate voltage.

Having established the topological-insulator behavior in the transistor mode, we now turn to the memristive configuration to explore how these edge states contribute to adaptive transport.

**Quantum spin Hall insulator in field-effect memristor configuration**

The realization of the topological memristor follows the operation scheme indicated in Fig. 1(d) and (e). In this configuration, the drain and top-gate voltages are shorted and sweeps of the bias voltage simultaneously control both the in-plane and perpendicular electric fields[34,35]. As was mentioned above, the back-gate offers additional control over the perpendicular electric field. We first consider the case where the band-gap region appears on the positive side for the down-sweep but at the negative side for the up-sweep, i.e. for back-gate voltages below ~ -8V.

Figure 3(a) shows the voltage-current characteristic of the memristor device when the bias current is swept between $\pm 10$ µA and the back-gate voltage is set constant to -10 V. The voltage is recorded as the voltage drop between the leads 5 and 4. A clear pinched hysteresis loop is observed between the down- and up-sweeps (black and red, respectively). For large positive currents, the in-plane and perpendicular electric fields are large, placing the Fermi energy in the conduction band, and resulting in n-type bulk conduction. Consequently, the voltage drops across the device and resistances are low. As the bias voltage is reduced towards zero, both in-plane and perpendicular electric fields decrease, pushing the Fermi energy into the topological band gap. This transition corresponds to the resistance peak observed near zero bias in the I-V curve. Further reduction of the bias voltage to -10 V increases the in-plane electric field and shifts the Fermi energy into the valence band, where transport is dominated by p-type bulk holes. Due to the charging of the floating gate for the up-sweep direction, the voltage-current characteristic shifts leading to the observed pinched hysteresis loop. Two key features characterize the transport behavior of our topological memristor. First, the transport mechanism depends on the sweep direction, as indicated by the color-coded boxes. At large in-plane electric fields, both sweep directions involve a single carrier type (the regions labeled with n/n and p/p). At lower fields however, the carrier type differs between sweeps (e.g. p-type for the up-sweep and n-type for the down-sweep). Finally, the regions labeled gap/n and gap/p are characterized by joint transport via the helical edge states and conductive bulk (n-type or p-type).

The second notable feature is the appearance of negative differential resistance (NDR) regions. Such NDR regions, also recently reported, in e.g., MOTT-Memristors, are of great interest for emulating neural functions using memristors[36,37]. Figs. 3(b) and 3(c) illustrate the two distinct transport regimes that define the functionality of the topological memristor. When the Fermi energy lies in the conduction or valence band, transport occurs through bulk states (n-type or p-type), leading to low-resistance values (Fig. 3(b)). This bulk resistance can be tuned by modulating the charge carrier density and mobility via the bias voltage, effectively enabling a variable conductance analogous to a tunable synaptic weight in a neural computation scenario. In contrast, when the Fermi energy is placed in the topological band gap, the conductivity through the bulk states is suppressed, and the transport is dominated by helical edge channels (Fig. 3(c)). This regime exhibits a stable resistance value representing a robust state, comparable to a fixed synaptic weight. Together, these two regimes offer the possibility to combine the flexibility of tunable bulk transport with the robustness of topologically protected edge conduction. However, it is important to note that the helical edge transport is not fully coherent over the entire device length, as the contact separation exceeds the phase coherence length[33,38]. To observe fully coherent and quantized edge transport, the contact separation must be reduced below the phase coherence length - a regime that will be explored in the next section (see Fig. 5).

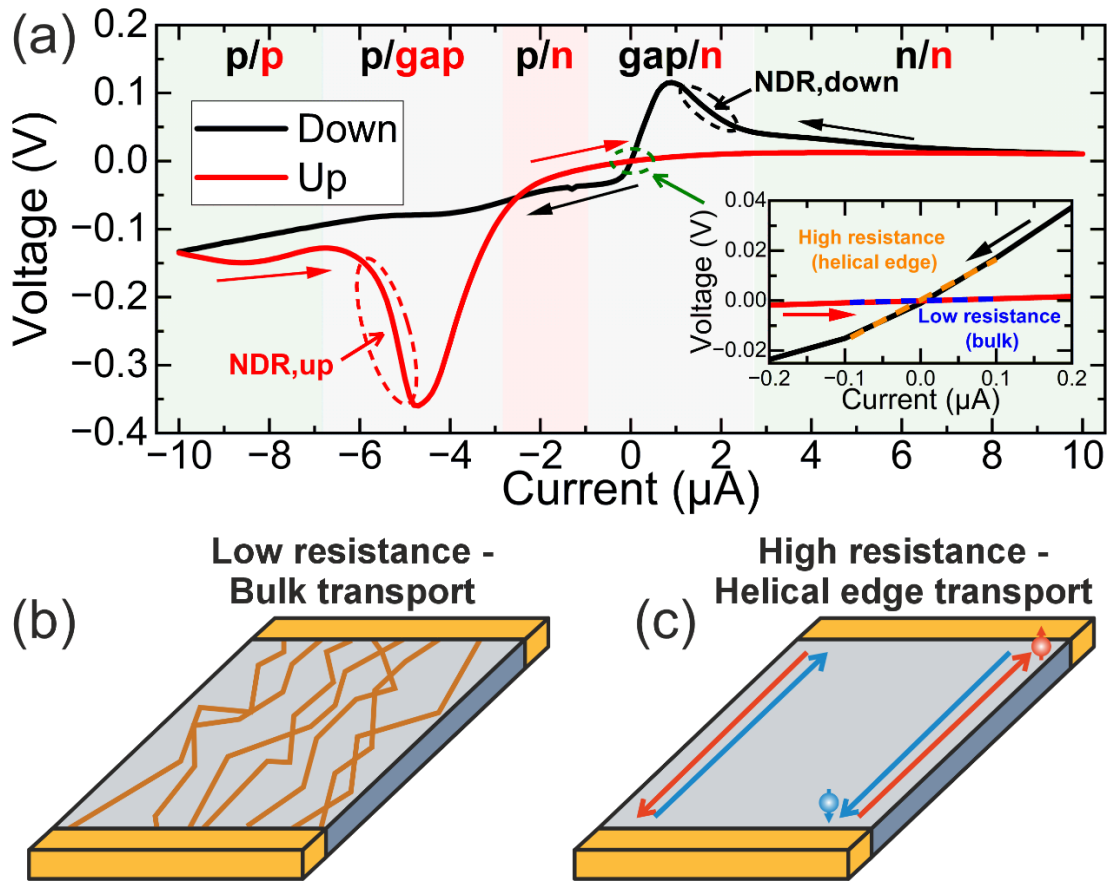

**Fig. 3. Transport features of the topological memristor.** (a) Voltage-current pinched hysteresis loop by sweeping the current between ±10 µA for a back-gate voltage of -10 V in black (down-sweep) and red (up-sweep), respectively. The device shows a clear pinched hysteresis loop with two resistance values emerging at zero bias current (High - helical edge and low - bulk, see inset). The color-coded boxes provide the regions with different combinations of carrier types responsible for transport dependent on the sweeping direction: p/p, n/n, p/n, gap/n and gap/p regions can be realized. (b),(c) Schematic of the two types of transport: low resistance via bulk transport in the n- and p-type regions and high resistance via helical edge transport in the topological gap regions.

To evaluate the performance of the topological memristor, we examine the change in resistance states under external electric fields applied, focusing on the tunability between high- and low-resistance states and their relative ratio. Figure 4 presents the detailed memristor response as a function of both the perpendicular (via back-gate voltage) and in-plane electric fields (via applied bias). Figure 4(a) shows a zoom-in of the voltage-current characteristics around zero bias voltage for two exemplary back-gate values: $V_{BG}$ = -2 V in black and $V_{BG}$ = -8V in blue. At $V_{BG}$ = -2 V, the slope of the voltage-current characteristics is almost flat and similar for both sweep directions, indicating a low resistance due to the bulk conductivity. The small difference in resistances mainly reflects variations in the charge carrier densities and mobilities resulting from different Fermi energy positions. In contrast, for $V_{BG}$ = -8 V, the down-sweep exhibits a much steeper voltage-current slope near zero bias than the up-sweep. This sharp increase in differential resistance indicates that the Fermi energy lies in the topological band gap during the down-sweep. Figure 4(b) depicts how the position of this maximum differential resistance evolves with back-gate voltage. Around $V_{BG}$ = -8 V, the resistance peak shifts towards zero bias, similar to the transistor behavior discussed previously (see Fig. 2(d)). This demonstrates that the perpendicular electric field can be used to align the topological band gap with the operating point of the memristor. Fig. 4(c) depicts the differential resistance value at zero bias for both the high-resistance (down-sweep) and low-resistance (up-sweep) states as a function of the back-gate voltage, for two different maximal bias voltages (±7 V and ±10 V). For a bias of ±10 V, the low-resistance state stays in the range below ~10 kΩ. In contrast, the high-resistance state exceeds 100 kΩ. The high-resistance state peaks near $V_{BG}$ = -8 V, where the topological band gap aligns with zero bias. In comparison, for a bias voltage of ±7 V, the difference between the high and low resistance states is less pronounced. The complete extracted High/Low ratios are summarized in Fig. 4(d). As shown above, the resistance ratio can be tuned by both the back-gate voltage and the bias amplitude. A maximum ratio of 22.5 is achieved for $V_{BG} \approx$ -8 V and ± 10 V bias, where the topological band-gap region fully overlaps with the high-resistance state at zero bias. This maximizes the contrast between the bulk and edge contributions to the conductivity.

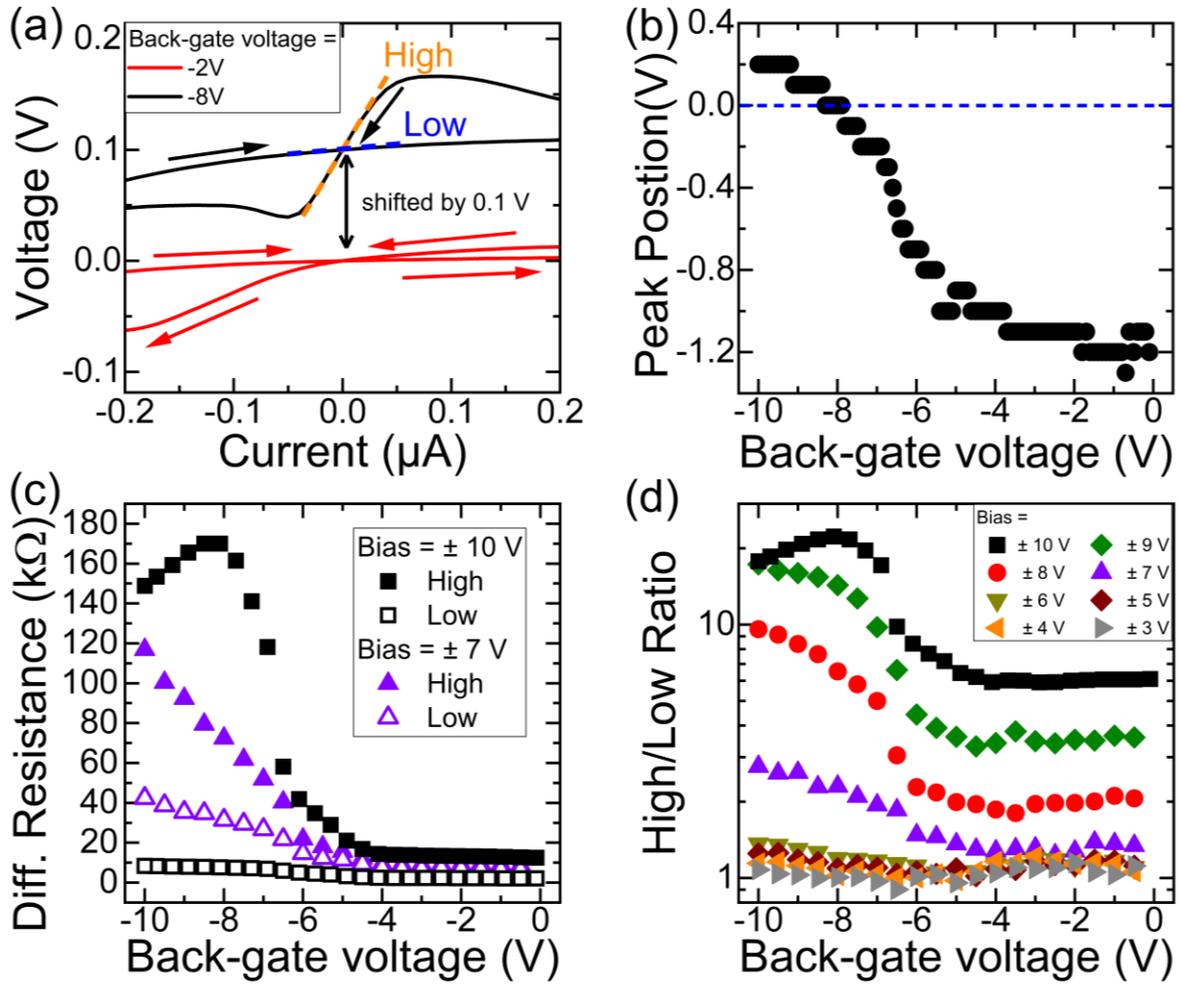

**Fig. 4. Electric field dependence of the topological memristor.** (a) Zoom-in of the voltage-current characteristics for back-gate voltages of -2 V (red) and -8 V (black) vertically offset by 0.1 V for clarity. The arrows indicate the sweeping direction. (b) Bias voltage position at the peak resistance in dependence of the back-gate voltage. At $V_{BG}$ = -8 V, the topological gap aligns with zero bias voltage. (c) Differential resistance at zero bias for the high- and low-resistance state as a function of back-gate voltage for two bias voltage amplitudes (±7 V in purple and ±10 V in black). (d) High/Low resistance ratios as a function of back-gate voltage for various bias voltage amplitudes between ±3 V and ±10 V. Below ±7 V, both resistance values almost coincide. At larger bias, the resistance ratio increases significantly and reaches a maximal value of 22.5 for a bias voltage of ±10 V and $V_{BG} \approx$ -8 V. This maximum is achieved when the topological gap coincides with zero bias voltage.

**Coherent helical edge transport in a topological memristor**

To directly probe coherent helical edge transport in the topological memristor, we fabricated a microscopic device with a contact separation length of 3 μm, which is shorter than the experimentally determined phase coherence length[30]. This microscopic design enables observation of memristive switching between a low-resistance incoherent bulk state and a high-resistance coherent helical edge state, highlighting the interplay between coherence and adaptive functionality in a single device. Figure 5(a) presents the voltage versus current (swept between ±10 μA) at zero back-gate voltage. The corresponding differential resistance (in units of $h/2e^2$) is shown in Fig. 5(b) for both current sweep directions (down-sweep in black, up-sweep in red). Strikingly, a robust resistance value emerges at $h/2e^2$ at around zero current, providing direct evidence of dissipationless quantized transport via topological helical edge states. This represents a clear manifestation of the QSHI state in the memristor configuration. While these measurements were performed at T = 4.2 K, the same material platform has exhibited quantized edge transport up to 60 K[30], indicating that the coherence underlying the memristive functionality can persist at elevated temperatures. During the up-sweep at zero current, the transport is dominated by incoherent bulk conductivity, resulting in a significantly lower differential resistance. This operating point of the memristor is indicated by the vertical dashed grey line. The differential resistance in the band-gap region during the up-sweep reaches approximately 1.3 $h/2e^2$, exceeding a quantized value because of the reduction in phase coherence length[33,39]. These results establish that memristive switching between a low-resistance of incoherent bulk states and a high-resistance of the coherent helical edge states can be realized in a single device. This demonstrates the successful integration of quantum-coherent and memory functionalities within a single topological field-effect memristor, highlighting its potential as a building block for hybrid quantum-neuromorphic architectures.

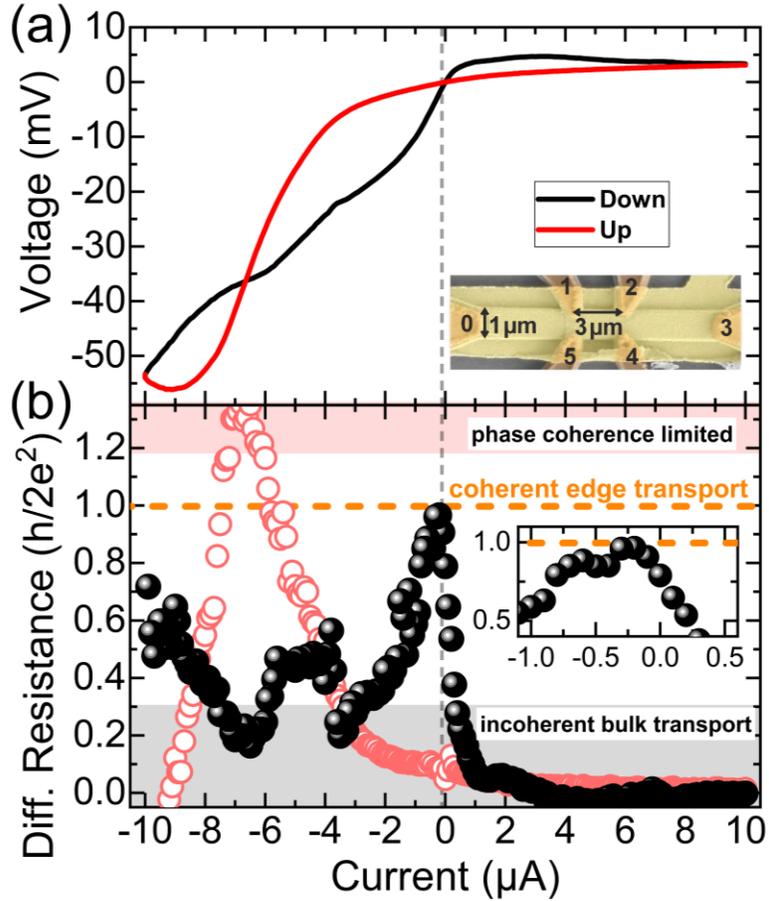

**Fig. 5. Microscopic topological memristor with coherent edge transport.** (a) Voltage as a function of current (down, black; up, red) in the memristor configuration for the microscopic device (see inset) with a contact separation length of L = 3 μm. (b) Differential resistance in units of $h/2e^2$, for the down- (black) and up-sweep direction (red). The pronounced quantized resistance value at $h/2e^2$ emerges due to dissipationless and coherent transport via helical edge states. For the up-sweep direction, the low resistance originates from incoherent bulk conduction.

**Conclusion**

In summary, we realized a topological field-effect memristor that combines quantized helical edge transport with memristive switching between distinct history-dependent transport regimes. Built on a scalable III–V semiconductor platform, the device exploits the helical edge states of InAs/GaInSb/InAs TQWs to switch between coherent topological and incoherent bulk conducting states. Crucially, the quantized resistance value of $h/2e^2$ in the microscopic device confirms the robustness and dissipationless character of the underlying topological edge states, even under device-level operation.

These results establish a direct link between topological protection and non-volatile memory functionality, introducing a new class of quantum-coherent memory elements. The dual transistor–memristor operation unifies topologically protected transport and adaptive charge-storage functionality within a single material system, enabling information to be stored and processed locally in the same coherent medium. This intrinsic coexistence of logic and memory offers a route to overcoming the von Neumann bottleneck. Beyond its conceptual implications, our work demonstrates a prototypical topological electronic device that combines coherence, adaptability, and scalability, opening the prospect of integrated quantum-neuromorphic hardware architectures.

**Acknowledgments**

The work was supported by the Elite Network of Bavaria within the graduate program "Topological Insulators" and by the Occitanie region through the TOP platform and the "Quantum Technologies Key Challenge" program (TARFEP project). We acknowledge financial support from the DFG within the project HO 5194/19-1 and through the Würzburg-Dresden Cluster of Excellence on Complexity and Topology in Quantum Matter – ct.qmat (EXC 2147, project-id 390858490). We also acknowledge the French Agence Nationale pour la Recherche (Cantor project - ANR-23-CE24-0022), and Equipex+ Hybat project - ANR-21-ESRE-0026), and the Physics Institutes of CNRS for Emergence 2024 - STEP - project. V. L.-R. acknowledges the support of CNPq Proj. 311536/2022-0 and FAPESP Projs. 2025/00677-8, and 2025/04805-0